\documentstyle[12pt]{article}
\def\be{\begin{equation}}
\def\ee{\end{equation}}
\begin{document}
\centerline{ \bf   On the history 
of fourth order metric theories of 
gravitation\footnote{Reprint of the 
original paper which appeared in
NTM-Schriftenr. Gesch. Naturw., Tech., Med. 
(Leipzig) {\bf 27} (1990) 1, pages 41-48; ISSN 0036-6978. 
The only difference in comparison with the original is that 
those text-parts which had been given only in German language, 
are now (in brackets just after the German text) translated into English.}}
\bigskip
\centerline{ By RAINER SCHIMMING (Greifswald)  
and HANS-J\"URGEN SCHMIDT (Potsdam)}
\medskip

Present address for correspondence: 

\centerline{http://www.physik.fu-berlin.de/\~{}hjschmi, \ 
e-mail: hjschmi@rz.uni-potsdam.de}
\medskip
\centerline{H.-J. Schmidt, Institut f\"ur Mathematik, Universit\"at
Potsdam}
\centerline{Am Neuen Palais 10,  D-14469 Potsdam, Germany}

\begin{abstract}
We present the history 
of fourth order metric theories of gravitation
from its beginning in 1918 until 1988. 
\end{abstract}

\section{Introduction}

From the advent of the general relativity theory (GRT) in 1915 by 
Albert Einstein (1879-1955) until today numerous geometrized theories 
of gravitation have been proposed.

Here, we shall review the history of a class of theories which is 
conceptually rather close to GRT:

\bigskip
\noindent 
- The gravitational field is described by a space-time metric only.

\bigskip
\noindent 
- The field equation follows from a Hamiltonian principle. The 
Lagrangian $L$ is a quadratic scalar in die Riemannian curvature of 
the metric. (Note that $L$ in GRT is linear in the curvature, i.e. 
proportional to the scalar curvature $R$.)

\bigskip
\noindent 
- The constants appearing in this ansatz are adjusted 
such that the 
theory is compatible with experimentally established facts.
Hence, the Lagrange function reads
\footnote{We apply the usual notations
 of tensor calculus and differential geometry. Particularly:
$R_{ijkl} = $ components of the 
curvature tensor of a Riemannian metric, 
$R_{ij} = $  components of the Ricci tensor $ = R^k_{\  ikj} $, 
$R =$ scalar curvature $= R^k_{\  k}$, 
$C_{ijkl} =$  components of the conformal curvature tensor.}
\be
L = aR^2 + b R_{ij}R^{ij} + kR + \Lambda
\ee
with constants $a, \  b, \  k, \  \Lambda$ where $a$ and $b$ do 
not vanish 
simultaneously.  The variational derivative of  
$R_{ijkl}  \,  R^{ijkl}  $  with respect to the metric 
can be linearly  expressed by the variational derivatives of   
$R_{ij} \, R^{ij}  $
and of  $R^2$  [1]. Thus we may omit  $R_{ijkl}  \,  R^{ijkl}  $
 in (1) without loss of  generality. 
The theory is scale-invariant if and
 only if   $ \Lambda \cdot k = 0$. 
It is even conformally invariant if and only if 
$\Lambda  = k = 0$ and $3a +b = 0$. 
The field equation following from $L$ eq. (1) is of fourth order, 
i.e.  
 it contains derivatives up to the fourth order of the components 
of the metric with respect to the space-time coordinates. 
(Note that Einstein's equation of GRT is of second order.)

The fourth order metric theories of gravitation are 
a very natural modification of the GRT. Historically, they 
have been introduced as a specialization of Hermann Weyl's 
(1885-1955) nonintegrable relativity theory from 1918 [2].  
Later on, just the fourth order theories became
interesting and more and more physical motivations supported 
them: The fourth order terms can prevent the big bang singularity 
of GRT; the gravitational potential of a point mass is 
bounded in the linearized case; the inflationary cosmological 
model is a natural outcome of this theory. But all the 
arguments from classical physics were not so convincing as 
those from quantum physics: the quantization of matter 
fields with unquantized gravity background leads to a 
gravitational Lagrangian of the above form [3]. Moreover, 
fourth
 order theories turned out to be renormalizable at the 
one-loop quantum level [4], but at the price of losing the 
unitarity of the S-matrix. (Note that Einstein's equation is not 
renormalizable.)  These circumstances caused a boom of fourth
 order gravity (classical as well as quantum) in the seventies. We will 
stop our record of the history before this boom. We restrict ourselves to 
the purely metrical theories (i.e., the affinity is always presumed 
to be Levi - Civita) and want only to mention here that fourth
 order field equations following from a variational principle
 can be formulated in scalar-tensor theories, theories with 
independent affinity, and other theories alternative to GRT
 as well.

\section{Papers inspired by Weyl's theory}

In 1918, soon after Albert Einstein's proclamation of the GRT,
 Hermann Weyl proposed a new kind of geometry and a
 unified theory of gravitation and electromagnetism based on
 it. He dwelled on the matter in a series of papers [2, 5-10] 
until it became superseded by the modern gauge field interpretation
 of electromagnetism [11 - 13]. Note that the gauge concept together
 with the words ``Eichung" (gauge) and ``Eichinvarianz" (gauge invariance) 
came 
into use in theoretical physics through Weyl's ansatz. For a 
broader 
discussion and evaluation we refer to [14]. A. Einstein [15] pointed 
out that the nonintegrability of the lengths of vectors 
under Weyl-like parallel propagation contradicts to physical 
experience. His argument has been refuted not earlier than 
in 1973: Paul Adrien Maurice Dirac (1902-1984) discusses 
the possibility of a varying gravitational constant. He writes:

``Such a variation would force one to modify  
Einstein's theory of gravitation. It is proposed that the 
modification should consist in the revival of Weyl's geometry, 
in which lengths are nonintegrable when carried around 
closed loops, the lack of integrability being connected 
with the electromagnetic field".  [16, p. 403]

H. Weyl's aesthetically very appealing modification of 
 GRT unfortunately does not directly
describe the real dynamics of fields and particles; 
however it deeply influenced the ``dynamics
of theories". By this we mean that various fundamental ideas 
have been formed or promoted by Weyl's papers:

\bigskip
\noindent 
- the search for alternatives to the GRT based on geometrization;

\bigskip
\noindent 
- the unification of the interactions or forces 
of nature, beginning with gravity and electromagnetism;

\bigskip
\noindent 
- field theories based on the geometry of an affine connection;

\bigskip
\noindent 
- conformal geometry and conformally invariant field theories;

\bigskip
\noindent 
- the gauge field idea, and

\bigskip
\noindent 
- fourth order gravitational field equations.

\bigskip

Here we are interested just in the last item. Weyl required 
the Lagrangian to be a polynomial function of the curvature 
and to be conformally invariant. He states:

``Dies hat zur Folge, dass unsere Theorie wohl 
auf die Maxwellschen elektromagnetischen nicht 
aber auf die Einsteinschen Gravitationsgleichungen 
f\"uhrt; an ihre Stelle treten Differentialgleichungen 4. 
Ordnung." [2, S. 477] (This has the consequence that
our theory, though it leads to Maxwell's equations  of
electromagnetism, fails to lead to Einstein's gravitational 
equations; they are replaced by differential equations of 
fourth order.)

The ambiguity in the concrete choice of  $L$ appeared as a 
difficulty which is opposed to the spirit of 
unification: any 
linear combination of  $R^2$ and  $R_{ij} \, R^{ij}  $
  would do. The variation of  $R_{ij} \, R^{ij}  $  or of
$R_{ijkl}  \,  R^{ijkl}  $  with respect to the vector field yields 
Maxwell-like equations, 
while for the choice of  $R^2$ an electromagnetic Lagrangian 
$F_{ij} \, F^{ij}  $
together with  a coupling constant $\alpha$ has to be added by 
hand:
$L = R^2 + \alpha  F_{ij} \, F^{ij}  $  [6, 2]. Weyl himself 
favoured different Lagrangians in different papers. 
Moreover,  he took trouble to produce results compatible 
with Einstein's GRT.  For this aim he destroyed the 
conformal invariance by a special 
gauge. Ernst Reichenb\"acher criticizes:

``Um so auffallender ist es, dass Weyl in dem von ihm
durchgerechneten Beispiel f\"ur die Wirkungsfunktion durch 
Festlegung der Eichung vor der Variation den Grundsatz 
der  Eichinvarianz durchbricht." [17, S. 157].
(It is even more conspicuous, that Weyl in his chosen
 calculated  example of an action has broken the axiom
of gauge invariance  before performing the variation.)

A more detailed analysis of the theory was necessary then. 
Roland Weitzenb\"ock [18] produced and studied all scalar invariants 
of the curvature in Weyl's geometry. Wolfgang Pauli jun. 
(1900-1958) [19, 20] and a little later Ferencz J\"uttner 
(geb.(born) 1878) [21] calculated the spherically symmetric 
static  gravitational field for variants of Weyl's theory. 
Pauli [20, S. 748]  comes to an important conclusion:

``Hiernach ist klar, dass aus Beobachtungen der 
Merkurperihelbewegung und der Strahlenablenkung, die
mit Einsteins Feldgleichungen im Einklang sind, niemals 
ein Argument gegen Weyls Theorie entnommen werden kann, 
wenigstens solange die letztere eine der
 drei Invarianten    $R_{ij} \, R^{ij}  $, $R^2$, 
$R_{ijkl}  \,  R^{ijkl}  $   
 als Weltfunktion zugrunde legt." (From this it becomes obvious,
that from observations of the Mercury perihelion change
and from the light ray deviation, which are in agreement with 
Einstein's field equations,  one can never deduce an argument
against Weyl's theory, 
at least, as long as one restricts to action functions
combined from the three invariants     $R_{ij} \, R^{ij}  $, $R^2$, 
$R_{ijkl}  \,  R^{ijkl}$.)

In other words, fourth order gravitational field equations 
following from (1) are not falsifiable by experimental physics! 
Pauli [20, 22] and other authors  did not even consider the vector 
field (i.e. assumed it to be equal to zero) in Weyl's theory, 
thus  making it unaffected by the criticism of nonintegrability [15]. 
Rudolf 
Bach [23] realized the possibility to keep the conformal invariance 
in a purely metrical theory: a Lagrangian 
$$
L = C_{ijkl}  \,  C^{ijkl} 
$$   
or   equivalently,
$$
L  = R^2 - 3R_{ij} \,  R^{ij}
$$
yields a conformally invariant field equation for the metric, later 
on called ``Bach's equation". In a similar spirit and in the 
same year 1921, Albert Einstein [24] proposed a 
conformally invariant theory. His expressions suffer from being 
non-rational in the metric. This theory is sometimes cited but 
has never been studied in details.

Reichenb\"acher [25, 17] proposed a variant of Weyl's theory based 
on a non-rational Lagrangian resembling nonlinear 
Born - Infeld electrodynamics.

In [26], also $L = R^2$ is used to get a field-theoretical 
model for the electron, but the fourth order terms are
 lost by an error in the calculations.

Cornel Lanczos (1893-1974) [27] tried a 
programme of ``Electromagnetismus als nat\"urliche 
Eigenschaft der Riemannschen Geometrie" (Electromagnetism 
as natural property of Riemannian geometry). 
He also assumed the vector field in Weyl's theory to be zero, but 
reintroduced 
it then in an alternative way as a set of Lagrangian multipliers. 
Unfortunately, Lanczos was, working 
with hyperbolic differential equations, misled by a formal analogy with
 elliptic differential equations. He varied the speculations with
 Lagrangian multipliers in a series of papers [28-34]. 
To take it positive, many useful mathematical formulas for fourth order 
theories 
resulted from Lanczos' work. Particularly, the paper [1] became 
a  ``citation classic".

In the twenties, the programme of classical field theory 
with its two cornerstones geometrization and unification lost 
some of its attractiveness in virtue of the quickly 
progressing quantum theory, cf. [35]. Moreover, there were the 
refutation of Weyl's theory and objections to fourth 
order equations. Lanczos expressed them as follows:

``Der Grund, weshalb diese Untersuchungen nicht weiter gediehen 
sind und zu keinem Fortschritt f\"uhrten, lag an zwei Momenten. 
Einerseits war es entmutigend, dass  man zumindest drei 
anscheinend gleichwertige Invarianten zur Verf\"ugung
 hatte: $R^2$, $R_ {\alpha  \beta}R^{\alpha \beta}$,
$R_ {\alpha \beta \gamma  \delta }R^{\alpha \beta  \gamma  \delta}$,
 ohne ein plausibles Auswahlprinzip zwischen
ihnen zu besitzen. Andererseits erscheinen diese Gleichungen, 
solange man ihre innere Struktur nicht verstehe, als 
Differentialgleichungen vierter Ordnung f\"ur  die die $g_{ik}$
von einer Kompliziertheit sind, die f\"ur jede weitere Schlussfolgerung 
ungeeignet ist." [27, p.75] (The reason, why these investigations 
did not give rise to further results, is twofold. On the one hand, it was
 discouraging, that one had at least three seemingly 
 similar invariants: $R^2$, $R_ {\alpha  \beta}R^{\alpha \beta}$,
$R_ {\alpha \beta \gamma  \delta }R^{\alpha \beta  \gamma  \delta}$,
without possessing any plausible principle of choice among them. 
On the other hand, these equations appear, as long as one does not
understand their inner structure, as differential equations of fourth 
order for the  $g_{ik}$ to be of such a complexity which makes them 
unsuited for drawing any further consequences from them.)

Similarly, Bergmann argues in his text-book [36] that, 
first, fourth order equations admit too many solutions  
and, second, their Lagrangian is rather ambiguous.

This situation explains why only  few papers on fourth 
order gravitation appeared in the period from the thirties to 
the sixties and why these did not follow the actual trends at 
their time. H. A. Buchdahl dealt with the subject in the period 
1948-1980. In his papers he covered the following problems: 

\bigskip 
\noindent 
- Invariant-theoretical considerations continuing those 
of Weitzenb\"ock and Lanczos [37];

\bigskip 
\noindent 
- General expressions for the variational derivatives of 
Lagrangians built from the curvature and, possibly, its 
derivatives are obtained [38-42].

\bigskip 
\noindent 
- Einstein spaces are solutions of a rather general class of 
fourth order equations [43, 44];

\bigskip 
\noindent 
- Static gravitational fields in fourth order theories [45];

\bigskip 
\noindent 
- Cosmological solutions in theories where the Lagrangian is 
a function of  the scalar curvature [46, 47];

\bigskip 
\noindent 
- Conformal gravity [48];

\bigskip 
\noindent 
- Reinterpretation of some fourth order equations in 
five dimensions [49].

\bigskip

Sir Arthur Stanley Eddington (1882-1944) in 1921, see [50], 
and Erwin Schr\"odinger (1887-1961) in 1948, see [51], 
also discussed gravitational field equations of fourth 
order to get field theoretical particle models, i.e., they 
tried to realize Einstein's particle programme.

\section{A new view}

Fourth order metric theories of gravitation have been 
discussed from 1918 up to now. One original motivation 
was the scale invariance of the action, a property which 
does not hold in GRT. Another motivation was the search 
for a unification of gravity with electromagnetism, which 
is only partially achieved with the Einstein-Maxwell system. 
There was no experimental fact contradicting GRT which could 
give motives for replacing it by a more complicated theory.

But a lot of problems appeared:

\bigskip 
\noindent 
1. The Lagrangian became ambiguous in sharp contrast to 
the required unification.

\bigskip 
\noindent 
2. The higher order of the field equation brought

\bigskip 
\noindent 
2.1. mathematical problems in the search for solutions and

\bigskip 
\noindent 
2.2. physical problems for the interpretation of the additional 
degrees of freedom.

\bigskip 
\noindent 
3. The well-founded Newtonian theory of gravitation did not 
result as the weak-field limit of
scale invariant fourth order gravity.

\bigskip

The third problem was the last of these to be realized but the 
first to be solved, both in 1947: One has to break the scale invariance
 of the theory by adding the Einstein-Hilbert action to the 
purely quadratic Lagrangian. Then, up to an exponentially 
small term, the correct Newtonian limit appears [52].

The original scale invariant theory then, again emerges as 
the high-energy limit of that sum. The items 1., 2. and 
the absence of experimental facts contradicting GRT seemed 
to restrain die research on these theories already in the twenties. 
Only in 1966 a renewed interest in these theories arose 
in connection with a semiclassical description of quantum 
gravity [53-55]. The coefficient of the quadratic term
 became calculable by a renormalization procedure, thus 
solving problem 1, at least concerning the vacuum equation. 
Further, the fact that fourth order gravity is one-loop 
renormalizable in contrast to GRT; a fact which was 
realized in 1977, [4] initiated a boom of research. 
It is interesting to observe that it is just the scale invariance 
of the curvature squared terms -- the original  motivation -- which 
is the reason for the renormalizability. Also the 
latest fundamental theory -- the superstring theory -- gives 
in the field theoretical limit (besides other terms) just 
a curvature-squared contribution to the action [56, 57]. The 
use of modern mathematics and computers has led 
to a lot of results to clarify the structure of the space 
of solutions thus solving problem 2.1. in the eighties. The 
more profound problem 2.2 has now three kinds of answers: 

\bigskip 
\noindent 
a) In spite of the higher order of the differential equation, a 
prescribed matter distribution plus the $O (1/r)$-behaviour of 
the gravitational potential suffice -- such as it takes place 
in Newtonian theory -- to determine the gravitational potential 
for isolated bodies in a unique way for the weak-field 
slow-motion limit, [52, 53]

\bigskip 
\noindent 
b) the observation that the additional degrees of freedom are 
just the phases of damped oscillations which become 
undetectably small during the cosmic evolution, and, by the 
way, can solve the missing mass problem and 
prevent the singularity problem of GRT [58], and

\bigskip 
\noindent 
c) it is supposed that there exist massive gravitons besides the 
usual massless gravitons known from GRT, but they 
are very weakly coupled [59].

\bigskip

The last point to be mentioned is the experimental testability: 
In the recent three years many
efforts have been made to increase the accuracy in determining 
the constants $G$, $\alpha$ and $l$ if the gravitational potential 
is assumed  (also in other theories than fourth 
order gravity) to be
$$
Gm r^{-1} (1 + \alpha e^{-r/l} ) \, . 
$$
The term proportional to  $\alpha$, the ``fifth force", can 
be interpreted as the fourth order correction to GRT. Up to 
now, it has not been possible to exclude  $\alpha =0$ by 
experiments [60-62].

Finally, let us say: Fourth order gravity theories will 
remain an essential link between GRT and quantum gravity 
for a long time.

\bigskip

{\bf   References }

[1]  Lanczos, K.: A remarkable property of the 
Riemann - Christoffel tensor in four dimensions. Annals of 
Math. 39 (1938) 842-850.

[2] Weyl, H.: Gravitation und Elektrizit\"at. Sitzungsber. 
Preuss. Akad. d. Wiss. Teil 1 (1918) 465-480.

[3] Utiyama. R., B. de Witt: Renormalization of a classical 
gravitational field interacting with quantized
matter fields, J. Math. Phys. 3 (1962) 608.

[4] Stelle, K.: Renormalization of higher-derivative 
quantum gravity. Phys. Rev. D 16 (1977) 953-969. 

[5] Weyl, H.: Reine Infinitesimalgeometrie. Mathemat. Zeitschr. 
2 (1918) 384-411.

[6] Weyl, H.: Eine neue Erweiterung der 
Relativit\"atstheorie. Ann. d. Phys. Leipz. (4) 59 (1919) 101-133. 

[7] Weyl, H.: Elektrizit\"at und Gravitation. 
Physik. Zeitschr. 21 (1920) 649-650.

[8]  Weyl, H.: \"Uber the physikalischen Grundlagen 
der erweiterten Relativit\"atstheorie. Physik. Zeitschr. 22 (1921) 473-480.

[9] Weyl, H.: Electricity and Gravitation. 
Nature 106 (1921) 800-802. 

[10] Weyl, H.: Raum, Zeit, Materie, 5. Auflage, Berlin: 
Springer-Verl. 1923.

[11] Schr\"odinger, E.: \"Uber eine bemerkenswerte 
Eigenschaft der Quantenbahnen eines einzelnen Elektrons. 
Zeitschr. f. Physik 12 (1923) 13-23.

[12] Weyl, H.: Elektron und Gravitation I. Zeitschr. f. 
Physik 56 (1929) 330-352.

[13] London, F.: Quantenmechanische Deutung der Theorie 
von Weyl, Zeitschr. f. Physik 42 (1927)
     375-389.  

[14] Vizgin, V. P.: Einstein, Hilbert, Weyl: 
Genesis des Programms der einheitlichen 
geometrischen Feldtheorien. NTM-Schriftenr. Leipzig 
21 (1984) 23-33.

[15] Einstein, A.: Nachtrag zu [2]; P. 478.

[16] Dirac, P. A. M.: Long range forces and broken symmetries. 
Proc. R. Soc. Lond. A 333 (1973) 403-418.

[17] Reichenb\"acher, E.: Die Eichinvarianz des Wirkungsintegrals 
und die Gestalt der Feldgleichungen in der
Weylschen Theorie. Z. f. Physik 22 (1924) 157-169.

[18] Weitzenb\"ock, R.: \"Uber die Wirkungsfunktion in der 
Weylschen Physik 1, 2, 3. Sitzungsber. Akad. d. 
Wiss. Wien, Math.-naturwiss. Kl. 
Abt. IIa, 129 (1920) 683-696; 697-703; 130 (1921) 15-23.

[19] Pauli, W.: Zur Theorie der Gravitation und 
der Elektrizit\"at von Hermann Weyl. Physik. Zeitschr. 
20 (1919) 457-467.

[20] Pauli, W.: Merkurperihelbewegung und 
Strahlenablenkung in Weyls Gravitationstheorie. 
Berichte d. Deutschen Phys. Ges. 21 (1919) 742-750.

[21] J\"uttner, F.: Beitr\"age zur Theorie der Materie. Math. 
Annalen 87 (1922) 270-306.

[22] Pauli, W.: Relativit\"atstheorie, Enc. math. Wiss. 
Bd. 5, Teil 2, S. 543-775. Leipzig: Teubner Verl. 1922.

[23] Bach, R.: Zur Weylschen Relativit\"atstheorie und 
der Weylschen Erweiterung des Kr\"ummungsbegriffs. 
Math. Zeitschr. 9 (1921) 110-135.

[24] Einstein. A.: Eine naheliegende Erg\"anzung des 
Fundamentes der allgemeinen Relativit\"atstheorie. 
Sitzungsbericht. Preuss. Akad. d. Wiss. Teil 1, (1921) 261-264.                      

[25] Reichenb\"acher, E.: Eine neue Erkl\"arung des 
Elektromagnetismus, Z. f. Physik 13 (1923) 221-240.

[26] Kakinuma, U.: On the structure of an electron, 
Part I, II. Proc. Phys.-Math. Soc. Japan Ser. 3, 
I  (1928) 235-242; II (1929) 1-11.

[27] Lanczos, C.: Elektromagnetismus als nat\"urliche 
Eigenschaft der Riemannschen Geometrie. Zeitschr. f. 
Physik 73 (1932) 147-168.

[28] Lanczos, C.: Zum Auftreten des Vektorpotentials in der 
Riemannschen Geometrie. Zeitschr. f. Physik 75 (1932) 63-77.

[29] Lanczos, C.: Electricity as a natural property of 
Riemannian geometry. Phys. Rev. 39 (1932)
716-736.

[30] Lanczos, C.: Ein neuer Aufbau der Weltgeometrie. 
Zeitschr. f. Physik 96 (1935) 76-106.

[31] Lanczos, C.: Matter waves and Electricity. Phys. Rev. 61 (1942) 
713-720.

[32] Lanczos, C.: Lagrangian multipliers and Riemannian spaces. 
Rev. Mod. Phys. 21 (1949) 497-502.

[33] Lanzcos, C.: Electricity and General Relativity. Rev. 
Mod. Phys. 29 (1957) 337-350.

[34] Lanczos, C.: Quadratic action principle of Relativity. 
J. Math. Phys. 10 (1969) 1057-1065.

[35] Vizgin V. P.: Hermann Weyl, die G\"ottinger 
Tradition der mathematischen Physik und einheitliche 
Feldtheorien. Wiss. Zeitschr. d. 
E.-M.-Arndt-Univ. Greifswald, Math.-naturwiss. Reihe 33  (1984)
57-60.

[36] Bergmann, P. G.: Introduction to the theory of relativity. 
New York: Prentice Hall 1942.

[37] Buchdahl, H.: On functionally constant invariants 
of the Riemann tensor. Proc. Cambr. Philos. Soc. 68 (1970) 179-185.

[38] Buchdahl, H.:  \"Uber die Variationsableitung
von Fundamentalinvarianten beliebig hoher Ordnung. Acta 
Mathematica 85 (1951) 63-72.

[39] Buchdahl, H.: On the Hamilton derivatives arising 
from a class of gauge-invariant action principles in a 
$W_n$. J. Lond. Math. Soc. 26 (1951) 139-149.

[40] Buchdahl, H.: An identity between the Hamiltonian 
derivatives of certain fundamental invariants in a $W_4$. 
J. Lond.  Math. Soc. 26 (1951) 150-152.

[41] Buchdahl, H.: On the gravitational field equations 
arising from the square of the Gaussian curvature. 
Nuovo Cim. 23 (1962) 141-156.

[42] Buchdahl, H.: The Hamiltonian derivatives of  a 
class of fundamental invariants. Quart. J. Math. Oxford
 19 (1948) 150.

[43] Buchdahl, H.: A special class of solutions of the 
equations of the gravitational field arising from 
certain gauge-invariant action principles. Proc. Nat. 
Acad. Sci. USA 34 (1948) 66-68.

[44] Buchdahl, H.: Reciprocal static metrics and 
non-linear Lagrangians. Tensor 21 (1970) 340-344.

[45] Buchdahl, H.: Quadratic Lagrangians and static gravitational 
fields. Proc. Cambr. Philos. Soc. 74 (1973) 145-148.

[46] Buchdahl, H.: Non-linear lagrangians and cosmological 
theory. Monthly Not. R. Astron. Soc. 150 (1970) 1-8.

[47] Buchdahl, H.: The field equations generated by the 
square of the scalar curvature: solutions of the Kasner type. 
J. Phys. A 11 (1978) 871-876.

[48] Buchdahl, H.: On a set of conform-invariant equations 
of the gravitational field. Proc. Edinburgh Math. Soc. 10  (1953) 
16-20.

[49] Buchdahl, H.: Remark on the equation 
$ \delta   R^2/ \delta g^{ij}  =0$.  Intern. 
J. Theor. Phys. 17 (1978) 149-151.

[50] Eddington, A. S.: Relativit\"atstheorie in mathematischer 
Behandlung. Berlin: Springer-Verl. 1925.

[51] Schr\"odinger, E.: Space-time structure. Cambridge: University 
Press 1950.

[52] Gregory, C.: Non-linear invariants and the problem of motion. 
Phys. Rev. 72  (1947) 72-75.

[53] Pechlaner, E., R. Sexl: On quadratic Lagrangians in 
General Relativity. Commun. Math. Phys. 2  (1966) 165-173.

[54] Sacharov, A. D.: Vakuumnye kvantovye fluktuacii v 
iskrivlennom prostranstve i teoria gravitacii. Dokl. Akad. Nauk 
SSSR 177 (1967) 70-71.

[55] Treder. H.-J.: Zur unitarisierten Gravitationstheorie mit 
lang- und kurzreichweitigen Termen. Ann. Phys. Leipz. 32 (1975) 
383-400.

[56] Ivanov, B.: Cosmological solution with string correction. 
Phys. Lett. B 198 (1987) 438.

[57] Hochberg, D., T. Shimada: Ambiguity in determining 
the effective action for string - corrected Einstein gravity. 
Progr. Theor. Phys. 78 (1987) 680.

[58] M\"uller, V., H.-J. Schmidt: On Bianchi type I vacuum 
solutions in $R + R^2$ theories of gravitation. I. The isotropic
 case.  Gen. Relat. Grav. 17 (1985) 769-781.

[59] Stelle, K.: Classical gravity with higher derivatives. Gen. 
Relativ. Grav. 9 (1978) 353-371. 

[60] Mio, N.: Experimental test of the law of gravitation at 
small distances. Phys. Rev. D 36 (1987) 2321.

[61] Stacey, F., G. Tuck, G. Moore: Quantum Gravity: 
Observational constraints on a pair of Yukawa terms.
Phys. Rev. D 36 (1987) 2374.

[62] Ander, M., M. Nieto: Possible resolution of the 
Brookhaven and E\"otv\"os experiments. Phys. Rev. Lett. 60 
(1988) 1225. 

\bigskip

{\bf Acknowledgement: } 
We thank Dr. sc. U. Kasper for critically reading the manuscript.

\bigskip

Anschriften der Verfasser: (Addresses of the authors in the year 1990:)

\bigskip
\noindent 
Dr. sc. mit. R. Schimming \\
Fachbereich Mathematik \\
Ernst-Moritz-Arndt-Universit\"at \\
F. -L.-Jahn-Str. 15a \\
Greifswald \\
DDR - 2200

\bigskip
\noindent 
Dr. sc. nat. H.-J. Schmidt \\
Zentralinstitut f\"ur Astrophysik \\
Akademie der Wissenschaften der DDR \\
R.-Luxemburg-Str. 17a \\
Potsdam \\
DDR - 1591

\end{document}